\documentclass[]{raa}
\usepackage{graphicx,times}
\usepackage{natbib}
\usepackage{amssymb,amsmath}
\usepackage{txfonts,booktabs,threeparttable}
\bibpunct{(}{)}{;}{a}{}{,}

\usepackage[a4paper=true,pagebackref=true]{hyperref}
\hypersetup{pdftitle = The title of my PDF, pdfauthor = My name, pdfsubject= The subject, pdfkeywords = keyword1 keyword2 keyword3}
\hypersetup{colorlinks = true, linkcolor = green, anchorcolor = red, citecolor = blue, filecolor = red, pagecolor = red, urlcolor = red}

\begin{document}

\title{The Fundamental Plane Relation of Early-Type Galaxies: Environmental Dependence}

   \volnopage{Vol.0 (200x) No.0, 000--000}      
   \setcounter{page}{1}          

   \author{Lei Hou
      \inst{1}
   \and Yu Wang
      \inst{1}
   }
+

   \institute{Department of Astronomy, University of Science and Technology of China, Jinzhai Road 96, Hefei 230026, China; {\it houleihl@mail.ustc.edu.cn}\\
   }

   \date{Received~~2009 month day; accepted~~2009~~month day}

\abstract{
Using a sample of 70,793 early-type galaxies from SDSS DR7, we study the environmental dependence of the fundamental plane relation. With the help of the galaxy group catalogue based on SDSS DR7, we calculate the fundamental planes in different dark matter halo mass bins for central and satellite galaxies respectively. We find the environmental dependence of the fundamental plane coefficients are similar in $g$, $r$, $i$ and $z$ bands. The environmental dependence for central and satellite galaxies is significantly different. While the fundamental plane coefficients of centrals vary systematically with the halo mass, those of satellites are similar in different halo mass bins. The discrepancy between centrals and satellites are significant in small halos, but negligible in the largest halo mass bins. These results remain the same when we only keep red galaxies, or galaxies with $b/a>0.6$, or galaxies in a specific radius range in the sample. After the correction of the sky background, results are still similar. We suggest that the different environmental effects of the halo mass on centrals and satellites may arise from the different quenching processes of them.
}

\authorrunning{L. Hou and Y. Wang}
\titlerunning{The Environmental Dependence of the Fundamental Plane}
\maketitle

\section{INTRODUCTION}
Galaxies obey several scaling relations between their dynamical and photometric properties. The Tully-Fisher relation \citep{1977A&A....54..661T} shows the brighter spiral galaxies prefer to have larger rotation velocities. For early-type galaxies (elliptical and lenticular galaxies, hereafter ETGs), velocity dispersion is correlated with both luminosity (the Faber-Jackson relation, \citet{1976ApJ...204..668F}) and the diameter in which the average surface brightness equals to a specified value (the $D_n-\sigma$ relation, \citet{1987ApJ...313...42D}). Moreover, there is a scaling relation between these 3 parameters: central velocity dispersion $\sigma_0$, effective radius $R_0$, and $I_0$, which is the average surface brightness in $R_0$ \citep{1987ApJ...313...59D,1987ApJ...313...42D}. This relation is called as the fundamental plane (hereafter FP), and is expressed as
\begin{equation}
\log R_0=a\log\sigma_0+b\log I_0+c.
\end{equation}
Both the Faber-Jackson relation and the $D_n-\sigma$ relation can be regarded as projections of the FP relation.

The FP relation reveals a lot of information about the dynamical properties of ETGs. In theory, the virial equilibrium expects the FP with $a=2$ and $b=-1$. However, the coefficients of the observed FP deviate from this result, which is called as the tilt of the FP. Moreover, it is remarkable that the uncertainty of the observed FP is very small, which amounts to the scatter in $R_o$ of about 20\%. Both the tilt and tightness of the FP provide strong constrains on the formation and evolution of ETGs.

Due to the tremendous development of galaxy redshift surveys in recent years, many studies focused on the FP relation of large samples. Some authors suggested that the FP is not an universal relation for all the ETGs but affected by the properties of ETGs. For example, coefficients and residuals of FP are correlated with luminosity, magnitude range, S\'ersic index, ellipticity, stellar mass-to-light ratio and color \citep[e.g.][]{2008ApJ...685..875D,2009MNRAS.392.1060N,2012MNRAS.427..245M,2013MNRAS.435...45D}. Some studies also found a systematic variation of stellar population parameters among the different positions on the FP \citep{2009ASPC..419...96G,2010MNRAS.408.1335L,2012MNRAS.420.2773S}.

Another interesting property of the FP is its environmental dependence. Although some studies declared there is no environmental dependence of the FP \citep{1996MNRAS.280..167J,2005MNRAS.360..693R}, others demonstrated the FP coefficients are significantly different between cluster and field ETGs, and between different clusters \citep{1991MNRAS.249..755L,1992ApJ...389L..49D,2013MNRAS.432.1709C}. The FP relation is also correlated with dark matter halo mass, group richness, cluster-centric distance, the distance to the Nth nearest neighbor and local galaxy density \citep{2003AJ....125.1866B,2006AJ....131.1288B,2008ApJ...685..875D,2010MNRAS.408.1361L,2012MNRAS.427..245M}. However, there is an important problem in these studies: different environmental indicators are adopted. As a result, these studies focused on different aspects of the environmental effects, and it is difficult to directly compare results of these studies.

In the standard $\Lambda$CDM cosmology, galaxies are born and located in dark matter halos. Photometric and dynamical properties of galaxies depend strongly on their host halo. The connection between galaxies and halos has been analysed with models such as the conditional luminosity function \citep{2003MNRAS.339.1057Y} and the halo occupation distribution \citep{1998ApJ...494....1J,2000MNRAS.318.1144P}. One popular environmental indicator is the mass of the host halo. For example, in larger halos galaxies prefer to be more massive and redder. Recently, many models prefer to distinguish whether the galaxy is a central or satellite one in the halo. Centrals are almost correlated to the halo mass, but satellites are affected by their accretion history \citep{2012ApJ...752...41Y,2013MNRAS.428.3121M}. The distinct between centrals and satellites indicates the position of the galaxy in the halo is another important environmental indicator.

In this work, we adopt the host halo mass and the position of the galaxy in the halo as environmental indicators, and investigate the environmental dependence of the FP relation. With the help of the group catalogue based on the Sloan Digital Sky Survey Data Release 7, hereafter SDSS DR7, we get the host halo and get the mass of the halo for each galaxy in this catalogue. Furthermore, we can recognize whether it is a central or satellite galaxy in the halo. Using these two environmental indicators, we investigate the correlation between the environment and the FP coefficients of SDSS ETGs.

The paper is organized as follows. Section 2 describes the ETG sample data from SDSS DR7, the calculation and corrections of the parameters, and the group catalogue which we use to characterize the environment of ETGs. In Section 3 we discuss the virial equilibrium of galaxies in dark matter halos, the fitting method of the FP, the results of the environmental dependence of the FP, and the test of systematics about the color and axial ratio. We summarize this paper and discuss our findings in Section 4. We adopt a $\Lambda$CDM cosmology with $\Omega_m=0.238$, $\Omega_\Lambda=0.762$ and $h=0.73$ in this work.

\section{GALAXY AND GROUP SAMPLES}
\subsection{Selecting the early-type galaxies}
This work is based on the data from SDSS DR7 \footnote{http://cas.sdss.org/dr7/en/}. We select ETGs using the following criteria, which is similar to those in \citet{2003AJ....125.1817B}:

(1) Concentration $r_{90}/r_{50}$ in $i$ band is larger than 2.5.

(2) Ratio of the likelihood of the de Vaucouleurs model to the exponential model $L_{deV}/L_{exp}\geq 1.03$.

(3) Spectral classification index $eClass$ is less than -0.1.

(4) Warning flag $zWarning$ is zero. This is the indicator of high spectral quality.

(5) Signal-to-noise ratio $S/N>10$.

(6) Redshift $z<0.2$. This criteria is for the purpose of coincidence with the redshift range of the group catalogue.

(7) Central velocity dispersion $\sigma>70~km~s^{-1}$. This is because the measure of $\sigma$ lower than this value has a significant uncertainty.

Number of galaxies in this ETG sample is 70,793.

To measure the FP, we should calculate the radius, the surface brightness and the velocity dispersion of these ETGs. The effective angular radius is
\begin{equation}
r_0=\sqrt{b/a}~r_{deV},
\end{equation}
where $b/a$ is the axis ratio, and $r_{deV}$ is the de Vaucouleurs angular radius. Given the redshift of this galaxy, we can convert the effective angular radius into the effective physical radius
\begin{equation}
R_0=r_0D_A(z),
\end{equation}
where $D_A(z)$ is the angular distance at the redshift $z$.

The mean surface brightness in $R_0$ is defined as $I_0=L/2R_0^2$. In this work, instead of measuring $I_0$ directly, we calculate the effective surface brightness $\mu_0\equiv-2.5\log_{10}I_0$, which is
\begin{equation}
\mu_0=m_{deV}+2.5\log_{10}(2\pi r_0^2)-K(z)-10\log_{10}(1+z)+Qz,
\end{equation}
where $m_{deV}$ is the extinction-corrected de Vaucouleurs magnitude, $K(z)$ is the K-correction to $z=0$, and $Q$ represents the correction factor for the evolution effect in luminosity. We use IDL code \emph{kcorrect} v4\_2 \footnote{http://howdy.physics.nyu.edu/index.php/Kcorrect} \citep{2007AJ....133..734B} to calculate $K(z)$. Values of evolution correction $Q$'s are taken from \citet{2003AJ....125.1849B}.

The central velocity dispersion $\sigma$ of a SDSS galaxy is estimated from the spectrum, which is observed using a fixed fiber aperture of 1.5 arcsec. As a result, $\sigma$ of a galaxy with a larger angular radius represents the motion of the more inner stars than another galaxy with a smaller radius \citep{1995MNRAS.276.1341J,1999MNRAS.305..259W}. We should do a aperture correction as
\begin{equation}
\sigma_0=\sigma(\frac{r_{fiber}}{r_0/8})^{0.04},
\end{equation}
where $r_{fiber}$ is the angular radius of the fiber, 1.5 arcsec, and $r_0$ is the effective radius calculated above.

\subsection{The SDSS group catalogue}
The group catalogue is constructed based on SDSS DR7 data using a modified version of the halo-based group finder developed in \citet{2005MNRAS.356.1293Y}. This catalogue is described in \citet{2007ApJ...671..153Y}. For each group in the catalogue, the dark matter halo mass $M_{halo}$ is estimated by two methods: $M_L$, estimation using the characteristic luminosity, and $M_S$, estimation using the characteristic stellar mass. In each group the brightest galaxy is recognized as the central galaxy, and the others are satellites of this group.

Making use of the SDSS group catalogue, for each galaxy in the ETG sample we find its host dark matter halo and the halo mass. Because $M_{halo}$ is more correlated to the stellar mass than to the luminosity, we adopt $M_S$ as the estimation of the halo mass in this work. It is recognized whether the galaxy is a central or satellite galaxy. Making use of the halo mass and the position in the halo as environmental parameters, we can study the environmental dependence of the FP relation in Section 3.3.

\section{THE FUNDAMENTAL PLANE RELATION OF EARLY-TYPE GALAXIES}
\subsection{Galaxies in virial equilibrium}
According to the virial theorem, galaxies in virial equilibrium should satisfy
\begin{equation}
\sigma^2\propto\frac{GM}{R}\propto \frac{M}{L}RI.\label{equat:virial}
\end{equation}
If all the ETGs have the same $M/L$, the FP should satisfy $a=2$ and $b=-1$. The tilt of the FP occurs when or $M/L$ is not constant but a function of $R$, $\sigma$ or $I$. The variation of  is $M/L$ due to the contribution of non-homology and stellar population.

Moreover, the dynamical properties of galaxies are remarkably affected by the dark matter halo. Firstly, the dark matter halo provide a potential well in which the stars and gas are located. In addition, tidal effects are exerted on satellite galaxies, which makes stars and gas of them more easily stripped. With the influence of the dark matter, the dynamical equilibrium of ETGs is
\begin{equation}
\sigma^2\propto\frac{GM}{R}+U_m,
\end{equation}
where $U_m$ represents the potential due to the dark matter halo. As a result, the tilt of the FP is correlated to dark matter halos.

Which of these factors plays the most important role in the tilt of the FP? It is a fairly controversial issue \citep{2004ApJ...611..739T,2005ApJ...623..666R,2006ApJ...649..599K,2006MNRAS.366.1126C,2006ApJ...640..662T,2007ApJ...665L.105B,2008ApJ...684..248B}. Recent studies suggest non-homology, stellar population and dark matter all contribute to the tilt, called as the "hybrid solution" \citep{2004ApJ...600L..39T,2009MNRAS.396.1171H,2013MNRAS.435...45D}.

\subsection{The FP relation}
Given the parameters $R_0$, $\mu_0$ and $\sigma_0$ calculated and corrected, we can fit the FP of the ETG sample. There are two methods to do this task. The direct fit is minimizing the scatter in the $R_0$ direction, while the orthogonal fit is minimizing the scatter in the orthogonal direction of the FP. Because the orthogonal fit treats the variables symmetrically, it is thought to be more physical than the direct fit. We adopt the orthogonal fit in this work.

First we calculate the covariance matrix of $I\equiv\log_{10}I_0$, $R\equiv\log_{10}R_0$, and $V\equiv\log_{10}\sigma_0$ of the ETG sample. We should subtract the error matrix from it to get the intrinsic covariance matrix (see the Appendix D of \citet{2003AJ....125.1817B} for details)
\begin{equation}
\mathcal{C} \equiv
\left( \begin{array}{ccc}
C_{II} & C_{IR} & C_{IV}\\
C_{IR} & C_{RR} & C_{RV}\\
C_{IV} & C_{RV} & C_{VV}
\end{array} \right).
\end{equation}
Then we diagonalize $\mathcal{C}$ and get the eigenvectors and eigenvalues. Alike to the principal component analysis, the eigenvectors represent the major axes of ETGs in $(I, R, V)$ space, and the eigenvalues are variances in the directions of these three axes. The eigenvector corresponding to the smallest eigenvalue is the normal vector of FP. We can derive FP slopes $a$ and $b$ from the normal vector. The intercept is determined as $c=\overline{R}-a\overline{V}-b\overline{I}$. The uncertainties of $a$, $b$ and $c$ are measured from 100-time bootstrap. FP coefficients of our ETG sample are shown in Table \ref{tab:FPetg}.

\begin{table}
\begin{center}
\caption[]{FP Fittings for our ETG Sample \label{tab:FPetg}}
 \begin{tabular}{cccccc}
  \hline\noalign{\smallskip}
Band & $a$ & $b$ & $c$ & $scatter_{orth}$ & $scatter_{R_0}$\\
  \hline\noalign{\smallskip}
g & $1.337\pm 0.004$ & $-0.751\pm 0.001$ & $-8.52\pm 0.01$ & 0.053 & 0.098\\
r & $1.373\pm 0.004$ & $-0.764\pm 0.001$ & $-8.45\pm 0.01$ & 0.051 & 0.096\\
i & $1.391\pm 0.004$ & $-0.776\pm 0.001$ & $-8.47\pm 0.01$ & 0.049 & 0.093\\
z & $1.427\pm 0.004$ & $-0.791\pm 0.002$ & $-8.57\pm 0.01$ & 0.049 & 0.093\\
  \noalign{\smallskip}\hline
\end{tabular}
\end{center}
\tablecomments{0.86\textwidth}{
FP fitting results in $g$, $r$, $i$ and $z$ bands of our ETG sample. Columns 2 to 4 are $a$, $b$, $c$. Columns 5 and 6 are scatters in the orthogonal direction and in the $R_0$ direction.
}
\end{table}

\subsection{Environmental dependence}
As is mentioned above, several definitions are used to describe the environment of galaxies, such as centric distance in the group, group richness and local galaxy density. In this work, we make use of the SDSS group catalogue, and adopt environmental indicators as the  mass of the halo and the position in the halo. The advantage of this definition is the adequate connection between galaxies and dark matter halos. In theory, galaxies formed from gaseous halos, which are remarkably correlated with the dark matter halo. In observations, color, morphology, 2-point correlation function, luminosity function and star formation history depend on the halo mass \citep[e.g.][]{2006MNRAS.366....2W,2012ApJ...752...41Y,2013MNRAS.428.3306W}. Moreover, dark matter halos affect centrals and satellites in distinct ways. Galaxies became satellites when they were accreted into a larger halo, with their star formation quenched easily due to strangulation. Meanwhile, the central galaxy is still accreting gas or merging with other galaxies \citep{2008MNRAS.387...79V,2013MNRAS.432..336W}. As a result, it is convenient to separate centrals and satellites when considering the environmental dependence.

Depending on the halo mass and the position in the halo, we divide the ETG sample into several subsamples. ETGs with halo mass between $10^{12.0}M_{\odot}$ and $10^{14.0}M_{\odot}$ are assigned into 8 halo mass bins. The size of each bin is 0.25 order of magnitude. ETGs in halos larger than $10^{14.0}M_{\odot}$ are set to a uniform bin. ETGs with halo mass lower than $10^{12.0}M_{\odot}$ are abandoned because the number of them is quite small and the uncertainty is large. In each bin, we separate ETGs as centrals and satellites. These subsamples are shown in Table \ref{tab:bin}.

\begin{table}
\begin{center}
\caption[]{Numbers of Centrals and Satellites in Each Halo Mass Bin \label{tab:bin}}
 \begin{tabular}{lll}
  \hline\noalign{\smallskip}
$\log M_{halo}$ & Centrals & Satellites\\
  \hline\noalign{\smallskip}
$(12.00, 12.25]$ & 9395 & 241\\
$(12.25, 12.50]$ & 9213 & 528\\
$(12.50, 12.75]$ & 8104 & 936\\
$(12.75, 13.00]$ & 5779 & 1396\\
$(13.00, 13.25]$ & 3848 & 1858\\
$(13.25, 13.50]$ & 2448 & 2345\\
$(13.50, 13.75]$ & 1459 & 2966\\
$(13.75, 14.00]$ & 742 & 2954\\
$(14.00, \infty)$ & 475 & 6565\\
  \noalign{\smallskip}\hline
\end{tabular}
\end{center}
\end{table}

\begin{figure}
\begin{center}
\includegraphics[scale=0.4]{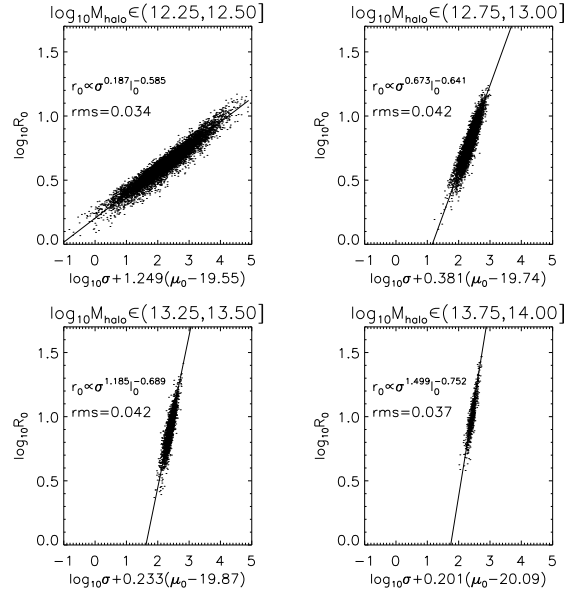}
\end{center}
\caption{Projected FPs of centrals in 4 halo mass bins in $r$ band. In each subsample, dots represent ETGs, and the slope of the solid line represents coefficient $a$ of the FP.}
\label{fig:FP_cen}
\end{figure}

\begin{figure}
\begin{center}
\includegraphics[scale=0.4]{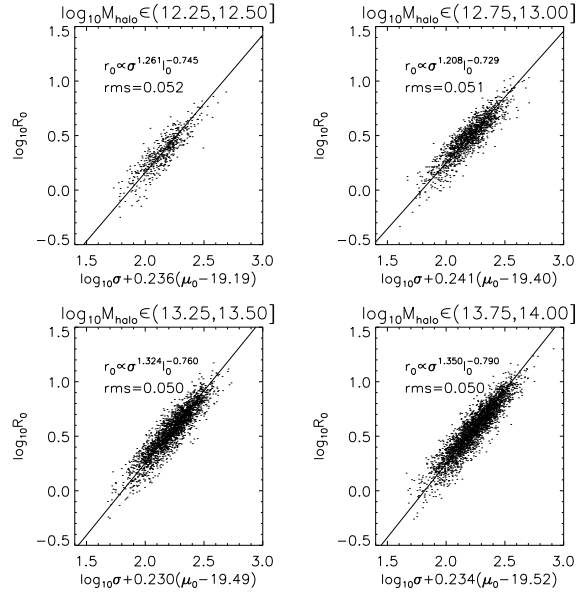}
\end{center}
\caption{Projected FPs of satellites in 4 halo mass bins in $r$ band. Symbols are similar to Figure \ref{fig:FP_cen}} \label{fig:FP_sat}
\end{figure}

\begin{figure}
\begin{center}
\includegraphics[scale=0.4]{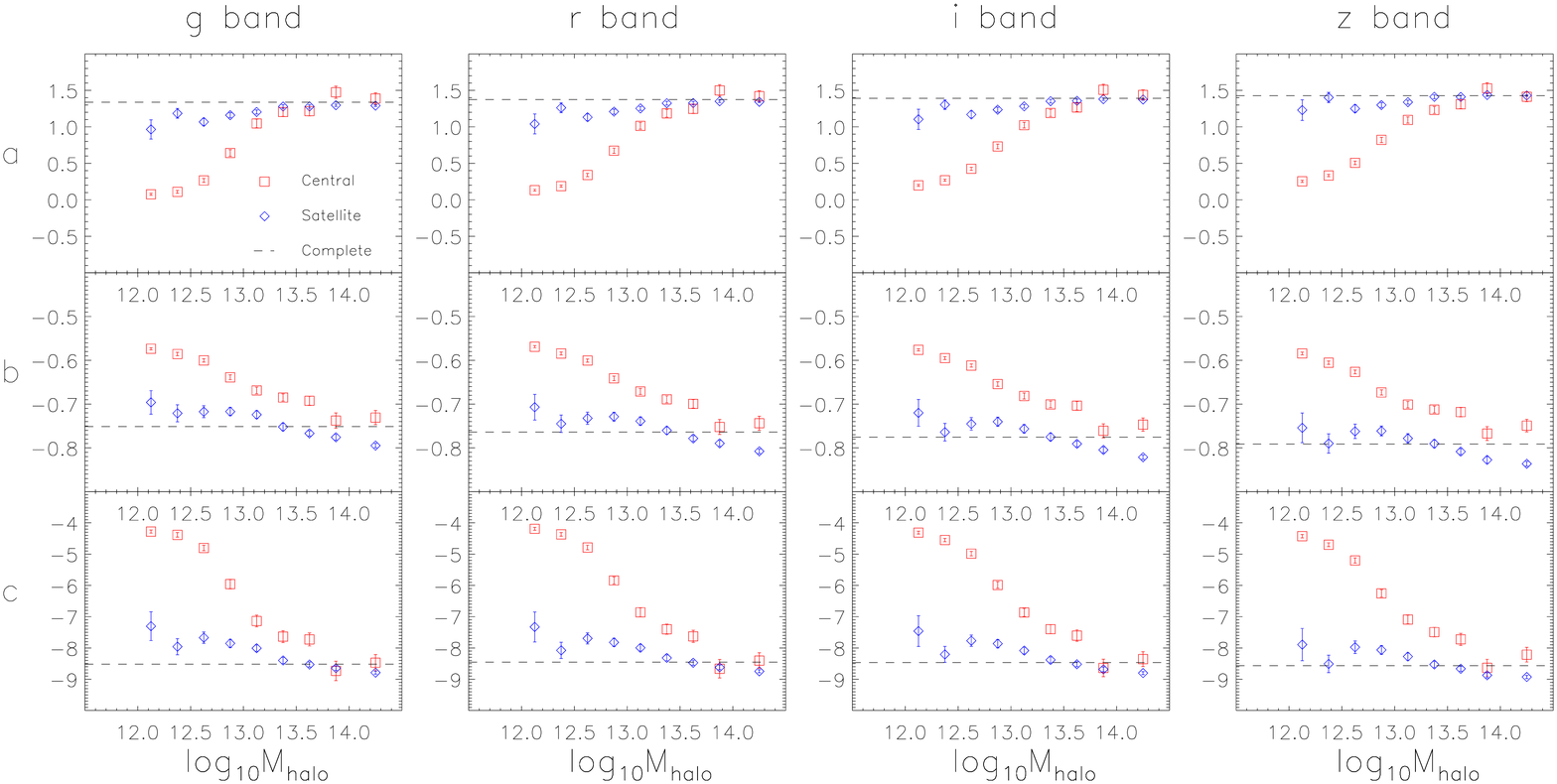}
\end{center}
\caption{The environmental dependence of the FP relation in 4 bands. Red squares are centrals, and blue diamonds are satellites. For each subsample, the value of $log_{10}M_{halo}$ is set as the midpoint of the corresponding halo mass bin. In the last halo mass bin $(14.00, \infty)$, $log_{10}M_{halo}$ is set as 14.25. The horizontal dashed lines are the corresponding coefficients of the complete ETG sample.} \label{fig:env_dep}
\end{figure}

FPs of these subsamples are calculated using the above fitting method. We find FPs of all the subsamples are well-shaped and tight. We show projected FPs of subsamples in Figures \ref{fig:FP_cen} and \ref{fig:FP_sat} for centrals and satellites respectively. In each figure, we only illustrate the shape of FPs in 4 halo mass bins as example. Units of $\sigma_0$, $\mu_0$ and $R_0$ are $km\,s^{-1}$, $mag\,arcsec^{-2}$ and $kpc$. We plot the environmental dependence of FP coefficients in Figure \ref{fig:env_dep}.

Several results are indicated in Figure \ref{fig:env_dep}. Firstly, FP coefficients as functions of the halo mass are similar in different bands. Secondly, all the FP coefficients of satellites are independent of the host halo mass, and are close to those of the complete ETG sample. Finally, for centrals, we find obvious correlations between FP coefficients and the halo mass. When the halo mass is increasing, $a$ is increasing, but $b$ and $c$ are decreasing. In small halos, FP coefficients of centrals are significantly different from those of satellites. In the halos which are more massive than $10^{13.25}M_{\odot}$, FP coefficients of centrals depend only weakly on the halo mass, and are close to those of satellites.

The distinction between the environmental dependence of centrals and satellites in Figure \ref{fig:env_dep} is noteworthy. This suggests that satellites are a relatively universal population but centrals are not. In small halos, there is a significant discrepancy between environment of centrals and satellites. In large halos, environment of centrals is similar to that of satellites.

\subsection{Systematics}
Selection criteria in Section 2.1 don't contain the color constraint or axial ratio constraint. There may be some late-type galaxies selected into our sample. The fraction of late-type galaxies increases with decreasing galaxy density, which means there is a larger proportion of late-type galaxies in small halos. We should test whether the environmental dependence of the FP is due to contamination of late-type galaxies.

\begin{figure}
\begin{center}
\includegraphics[scale=0.4]{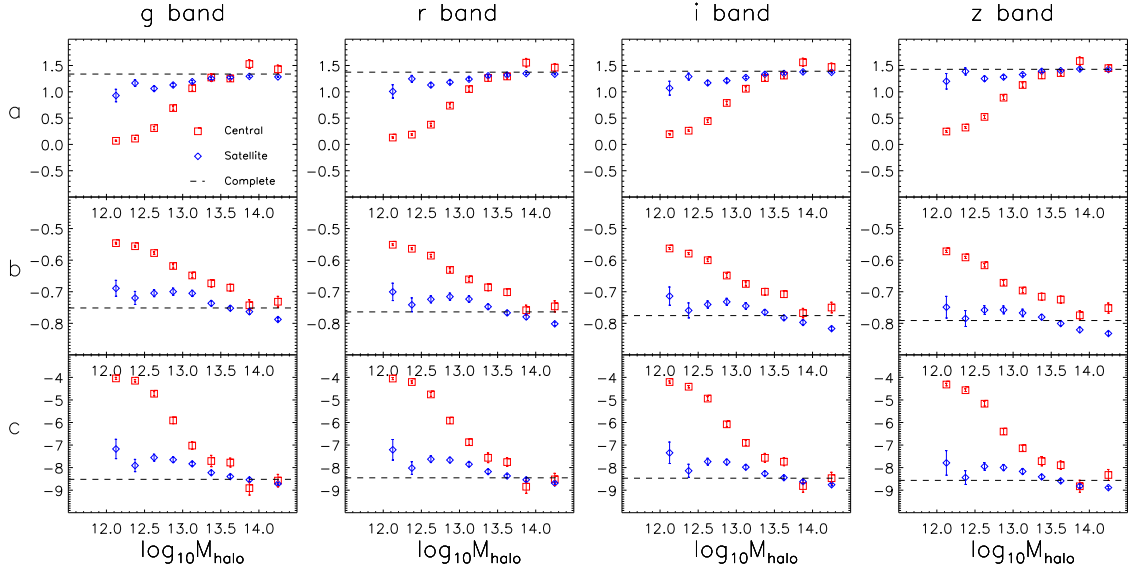}
\end{center}
\caption{Environmental dependence of FP coefficients of red ETGs. Symbols are similar to Figure \ref{fig:env_dep}.} \label{fig:env_red}
\end{figure}

To do this, all the galaxies in the SDSS DR7 are divided into 120 luminosity bins in such a way that in each bin there are the same number of galaxies. In each bin, we fit the $g-r$ color distribution of galaxies to the double Gaussian, get the $g-r$ values of the red peak and the blue peak, and calculate the average color of these two peaks. Then a linear fitting is done between the average colors and the average magnitudes in these bins. We get the fitting result as $g-r=0.68-0.030(M_r +21)$, and adopt this as the dividing line of red and blue galaxies: galaxies above this line are selected as red ones, others are blue ones. For each subsample in Table \ref{tab:bin}, we abandon blue galaxies to get a corresponding red subsample, and measure its FP relation. FP coefficients of red subsamples are plotted in Figure \ref{fig:env_red}. We find the dependence of red ETGs are akin to Figure \ref{fig:env_dep}.

\begin{figure}
\begin{center}
\includegraphics[scale=0.4]{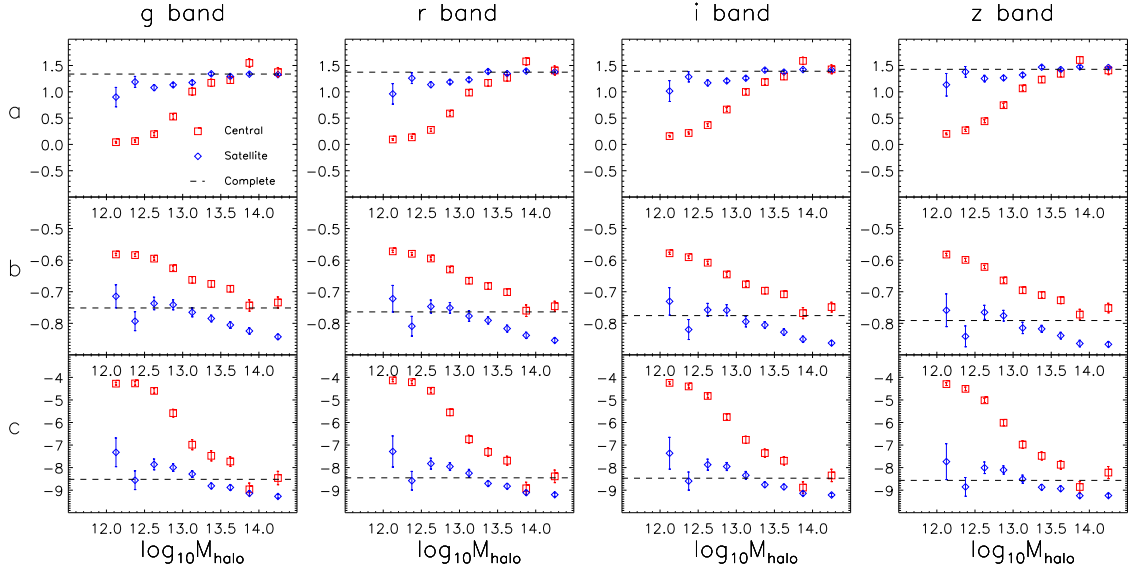}
\end{center}
\caption{Environmental dependence of FP coefficients of ETGs with $b/a>0.6$. Symbols are similar to Figure \ref{fig:env_dep}.} \label{fig:env_ar}
\end{figure}

Similarly, we can select early-type galaxies with axial selection criterion. For each subsample in Table \ref{tab:bin}, we only keep the galaxies with axial ratio $b/a>0.6$, and measure the FP of the remaining galaxies. As is shown in Figure \ref{fig:env_ar}, the environmental dependence of FPs for these galaxies is also akin to Figure \ref{fig:env_dep}.

For a given halo mass bin, centrals and satellites have different stellar mass distribution, especially for the low halo mass bins \citep{2012ApJ...752...41Y}. This may bias the FP fitting. Many works indicated that there is a correlation between size and the stellar mass of galaxies \citep{2006MNRAS.373L..36T,2008A&A...482...21C,2008ApJ...677L...5V,2012MNRAS.427.1666B,2014MNRAS.443..874B}. Therefore, if we force centrals and satellites with the similar $\log R_0$ distribution, in the first order, galaxies would have similar stellar mass distrubition. This ensures that the FP fittings are compared in a fair and unbiased way. To do this, we fit the $\log R_0$ distribution of ETGs in each halo mass bin to the double Gaussian. Assuming the fitted peaks and standard deviations are $\mu_1$, $\sigma_1$, $\mu_2$ and $\sigma_1$, with $\mu_1<\mu2$, we only keep the galaxies with $\log R_0$ in the range [$\mu_1$-$\sigma_1$, $\mu_2$+$\sigma_1$], and calculate the environmental dependence of the remaining subsamples. We find the result is consistent with Figure \ref{fig:env_dep}.

In a word, after the color selection, axial ratio selection and radius selection, our results in Figure \ref{fig:env_dep} still remain. This proves the environmental dependence of the FP relation is not due to the containment of late-type galaxies.

Recently, several works indicated that the sky subtraction algorithm of SDSS DR7 systematically overestimates the sky background of large galaxies and galaxies in dense regions. Therefore, the surface brightness and the half-light radii of centrals are underestimated \citep{2007AJ....133.1741B,2007ApJ...662..808L,2007MNRAS.379..867V,2013ApJ...773...37H,2011ApJS..193...29A}. It is suggested that the sky background should be estimated as the global sky of the field rather than the local sky which is adopted in the SDSS pipeline. To test whether this results in a bias in the FP, we corrected this bias as 
\begin{equation}
I_{0corr}=I_0+I_{LocalSky}-I_{GlobalSky},
\end{equation}
and re-fit the FPs. FP fitting results of the ETG sample with sky background correction are shown in Table \ref{tab:sky}, which are similar to those in Table \ref{tab:FPetg}. We also find that the sky background correction does not affect the environmental dependence of FP coefficients. Moreover, we also analyse the effect of the bias on the galaxy size. The effective sizes of centrals are underestimated, and this bias is more significant for larger centrals \citep{2011ApJS..193...29A}. Imagining that the $\log R_0$ is more underestimated at the large $\log R_0$ end of the FP, we can qualitatively find this makes coefficient $a$ lower, and coefficient $b$ higher. Meanwhile, this bias is more significant for centrals in larger halos. Therefore, for centrals in larger halos, coefficient $a$ should get more positive correction and coefficient $b$ should get more negative correction. This would strengthen the environment dependence of coefficients $a$ and $b$, rather than weaken the dependence. Therefore, the bias on the surface brightness and the effective size would not affect our results.

\begin{table}
\begin{center}
\caption[]{Fittings for our ETG Sample with Sky Background Correction \label{tab:sky}}
 \begin{tabular}{cccccc}
  \hline\noalign{\smallskip}
Band & $a$ & $b$ & $c$ & $scatter_{orth}$ & $scatter_{R_0}$\\
  \hline\noalign{\smallskip}
g & $1.340\pm 0.004$ & $-0.737\pm 0.002$ & $-8.55\pm 0.01$ & 0.053 & 0.098\\
r & $1.375\pm 0.004$ & $-0.752\pm 0.002$ & $-8.51\pm 0.01$ & 0.052 & 0.096\\
i & $1.392\pm 0.003$ & $-0.768\pm 0.002$ & $-8.55\pm 0.01$ & 0.050 & 0.093\\
z & $1.429\pm 0.003$ & $-0.781\pm 0.002$ & $-8.64\pm 0.01$ & 0.049 & 0.093\\
  \noalign{\smallskip}\hline
\end{tabular}
\end{center}
\tablecomments{0.86\textwidth}{
FP fitting results of the ETG sample with sky background correction. Columns are similar as Table \ref{tab:FPetg}.
}
\end{table}

\section{DISCUSSION AND CONCLUSION}
In this paper we make use of the data from SDSS DR7 to study the environmental dependence of the FP relation of 70,793 ETGs. Due to the SDSS group catalogue, for each galaxy we get the mass of its host dark matter halo, and specify it as a central or satellite galaxy in the halo. We investigate how FP coefficients depend on the halo mass and the position in the halo. We find the main results as follows.

(1) The environmental dependence of the FP relation is similar in $g$, $r$, $i$ and $z$ bands.

(2) FP coefficients of satellites are independent of their host halo mass, and are close to those of the complete ETG sample.

(2) FP coefficients of centrals show significant dependence on the halo mass. We find $b$ and $c$ decrease but $a$ increase with the halo mass. In small halos, the discrepancy between centrals and satellites is significant. In the largest halos, FP coefficients are similar to those of satellites.

(4) These relations still remain even when we only keep the red galaxies, or galaxies with $b/a > 0.6$, or galaxies with radius in a specific range.
Moreover, the sky background correction does not affect these results.

There are several studies on the correlation between FP coefficients and the environmental density.  \citet{2008ApJ...685..875D} studied the WIde-field Nearby Galaxy-cluster Survey (WINGS) sample, and found in denser environment $a$ is larger but $b$ and $c$ are smaller. However, \citet{2010MNRAS.408.1361L} combined SDSS and UKIDSS data and got a different result. They found although the dependence of $b$ and $c$ is similar to \citet{2008ApJ...685..875D}, $a$ is smaller in denser region. They explained that the discrepancy with \citet{2008ApJ...685..875D} might result from the fact that they corrected two biases of the FP slopes: one is due to the variation of averaged $M/L$ from field to group galaxies, and the other is that the field and group have the different distribution in the parameter space.  Moreover, \citet{2012MNRAS.427..245M} investigated about 10,000 ETGs in the 6dF Galaxy Survey. They found $a$ and $b$ are independent on the environment, and  $c$ is smaller in denser environment.  All these studies found the global environmental dependence (cluster-centric distance, local galaxy density, etc.) of the FP is similar to the local environmental dependence (dark matter halo mass, group richness, etc.) but weaker.

All these studies did not distinguish between centrals and satellites as in this paper. For satellites we did not find the correlation between FP coefficients and the halo mass. For centrals, our findings is consistent with \citet{2008ApJ...685..875D}, but conflicts with the dependence of $a$ in \citet{2010MNRAS.408.1361L}. Because \citet{2010MNRAS.408.1361L} did two corrections, and because the correlation between $a$ and the global environmental parameters is only significant at about $2\sigma$, this is not a severe conflict.

As is mentioned above, the tilt of the FP is the result of the non-constant $M/L$. This is due to non-homology and/or the variation of stellar population. 
One example of non-homology is the density profile of the dark matter halo. Centrals live in the centre of the halo, where the dark matter density is higher and more sensitive to the halo mass than where satellites live. This may be one explanation of the environmental dependence found in this work. Moreover, when fitting the FP, luminosity $L$ rather than stellar mass $M_*$ is used. This introduces the contribution of stellar population (such as $M_*/L$ or color) into the FP. Indeed, we find the environmental dependence of color is different for centrals and satellites. The color of centrals is redder in larger halos, but the color of satellites depends only weakly on the halo mass. \citet{2009MNRAS.396.1171H} indicated color can be treated as the fourth parameter of the FP. To explore these, the environmental dependence of the stellar mass FP should be investigated. In the future, we will come back to this and study the contribution of stellar population to the variation of FP coefficients.

The distinct between the trends of centrals and satellites on the halo mass demonstrates the dark matter halo affects centrals and satellites in different ways. Indeed, several studies found the quenching process of satellites is remarkably different from centrals. \citet{2012ApJ...757....4P} found the fraction of quenched centrals depends on the halo mass, but the quenched faction and the star formation rate (SFR) of satellites is mainly driven by the local density and independent on the halo mass. \citet{2013MNRAS.432..336W} proved the star formation rate (SFR) of centrals in larger halos peaks at a larger redshift; for satellites, the star formation evolved similar as centrals before they fell into the main halo, afterward it is quenched rapidly in a time of 0.2-0.8 Gyr; this quenching time-scale is independent of the halo mass. \citet{2013MNRAS.428.3121M} also investigated concluded the star formation rate of centrals depends on the halo mass, and the stellar mass of satellites is determined by the mass of its subhalo when it is falling into the main halo.

All these studies suggest the quenching process of centrals is mainly determined by the halo mass, that is, the global environment, but the quenching of satellites depends on the local environment, which is little correlated with the mass of the main halo. This is consistent with results in this paper. Furthermore, the local density of the centre of a halo is correlated with the halo mass. That means the quenching process of centrals depends on both the global and local environment, but satellites is only correlated with the local environment. This explains why the global environmental dependence of the FP is weaker than the local environmental dependence. In conclusion, the distinguish between the environmental dependence of the FP relation for centrals and satellites may be driven by the different quenching processes of them.

\begin{acknowledgements}
We are grateful to the Xiaohu Yang for the SDSS group catalogue. WY like to acknowledge the support of the Fundamental Research Funds for the Central Universities.

Funding for  the SDSS and SDSS-II has  been provided by the  Alfred P. Sloan Foundation, the Participating Institutions, the National Science Foundation, the  U.S.  Department of Energy,  the National Aeronautics and Space Administration, the  Japanese Monbukagakusho, the Max Planck Society, and  the Higher Education  Funding Council for  England.  The SDSS Web  Site is  http://www.sdss.org/.  The SDSS  is managed  by the Astrophysical Research Consortium  for the Participating Institutions. The  Participating Institutions  are the American  Museum of  Natural History,  Astrophysical   Institute Potsdam,  University   of  Basel, Cambridge University,  Case Western Reserve  University, University of Chicago,  Drexel University,  Fermilab,  the Institute  for  Advanced Study, the Japan Participation  Group, Johns Hopkins  University, the Joint Institute for  Nuclear  Astrophysics, the  Kavli Institute  for Particle Astrophysics  and Cosmology, the Korean  Scientist Group, the Chinese Academy of Sciences  (LAMOST), Los Alamos National Laboratory, the Max-Planck-Institute for Astronomy (MPIA),     the Max-Planck-Institute   for  Astrophysics   (MPA), New   Mexico  State University,   Ohio  State   University, University  of   Pittsburgh, University  of  Portsmouth, Princeton University,  the United  States Naval Observatory, and the University of Washington.
\end{acknowledgements}

\bibliographystyle{raa}
\bibliography{fp}

\end{document}